\documentstyle [12pt,dina4,epsf]{article}
\begin{document}
\parindent=0cm
\parskip=1.5mm

\def\bi{\begin{list}{$\bullet$}{\parsep=0.5\baselineskip
\topsep=\parsep \itemsep=0pt}}
\def\ei{\end{list}}
\def\phi{\varphi}
\def\-{{\bf --}}
\def\vm{v_{max}}

\begin{center}
{\LARGE\bf Cellular automaton models}
\end{center}
\begin{center}
{\LARGE\bf and traffic flow }
\end{center}
\vskip1.8cm
\setcounter{footnote}{1}
\begin{center}
{\Large A.\ Schadschneider\footnote{email: \tt as@thp.uni-koeln.de} and
M.\ Schreckenberg\footnote{email: \tt schreck@thp.uni-koeln.de} }
\end{center}
\vskip1.3cm
\begin{center}
Institut f\"ur Theoretische Physik\\
Universit\"at zu K\"oln\\
D--50937 K\"oln, F.R.G.
\end{center}
\vskip1.3cm
\vskip2cm
{\large \bf Abstract}\hspace{.4cm} A recently introduced cellular
automaton model for the description of traffic flow is investigated.
It generalises asymmetric exclusion models which have attracted
a lot of interest in the past.
We calculate the so-called fundamental diagram (flow vs.\ density)
for parallel dynamics using an improved mean-field approximation which
takes into account short-range correlations. For maximum velocity 1
we find  that the  simplest non-trivial of these approximations gives
already the exact result. For higher velocities our results are in
excellent agreement with numerical data.
\vfill
\pagebreak
The investigation of traffic flow in the past was based mainly on the
use of fluid-dynamical methods. In recent years also methods of
nonlinear dynamics were applied. On the other hand, due to their
computational simplicity, lattice gas automata \cite{Wo} were
succesfully applied to simulate fluids \cite{FHP86} and traffic in
one \cite{NS92} and two dimensions \cite{BML,CMMS}.  Similar models
have also been used for the description of asymmetric exclusion
processes \cite{KDN}-\cite{SCH93} and surface roughening
\cite{TANG92}.  Several exact solutions have been obtained for
asymmetric exclusion processes where the particles can move at most
one lattice spacing per update step. In a more general situation one
considers particles which can move over larger distances. All these
models may be interpreted as discrete models for the simulation of
traffic flow. Starting from realistic traffic one usually has a whole
spectrum of car velocities and thus it is straightforward to
introduce these more general exclusion models which then are more
appropriate for comparison with 'experiments' (i.e.~measurements on
freeway traffic
\cite{HBG86}).

In this letter we will consider single-lane traffic on a ring
geometry (length $L$, periodic boundary conditions). Generalizations
to multi-lane traffic and other boundary conditions (e.g.\ the
bottleneck situation where one considers open boundaries) will also
be discussed briefly.

The exact definiton of the model following \cite{NS92} is as
follows:\\
On a ring with $L$ sites every site can either be empty or
occupied by one vehicle with velocity $v=0,1,\ldots ,\vm$. At each
discrete time-step $t\to t+1$ an arbitary arrangement of $N$ cars is
updated according to the following rules:
\bi

\item[1)]
{\bf Acceleration:} If the velocity~$v$ of a vehicle is lower than
$\vm$ the speed is advanced by one [$v_1=v+1$].

\item[2)] {\bf Slowing down (due to other cars):}
If the distance $d$ to the next car ahead is not larger than $v_1$ ($d \le
v_1$)
the speed is reduced to $d-1$ [$v_2 = d-1$].

\item[3)] {\bf Randomization:}
With probability $p$, the velocity of a vehicle
(if greater than zero) is decreased by one [$v_3=v_2-1$].

\item[4)] {\bf Car motion:}
Each vehicle is advanced $v=v_3$ sites.

\ei

These rules can be applied to all cars in parallel (parallel update),
sequentially to randomly chosen cars (random-sequential update) or in
parallel to all cars on even and odd lattice sites
(sublattice-update)\footnote{Often the sublattice-update is also
called parallel update.}.  The rules ensure that the total number $N$
of cars is conserved under the dynamics (this is not true in the
bottleneck-situation). Note that even for parallel update the
randomization yields non-deterministic behaviour. For
random-sequential update the probability $p>0$ is not essential
because it only rescales the time axis \cite{NS92}. In the following
we will concentrate on the cases $\vm=1,2$ and parallel update.

In the simplest case $\vm=1$ the cars are allowed to move only one
step during an update. For this situation several results are known
\cite{KDN}-\cite{SCH93}.  Especially it can be shown that for
random-sequential update the mean-field Ansatz yields the exact
equilibrium state \cite{NS92,GS92}, which is equivalent to the fact
that for a fixed number of cars every arrangement of cars occurs with
the same probability. Therefore it is quite natural to take the
mean-field approach also as a starting point for the investigation of
higher velocities $\vm>1$ and parallel update.

Our main interest will be the calculation of the so-called
fundamental diagram (flow $q$ vs.\ density $\rho =N/L$). As described
in \cite{NS92} these results can be compared directly with
measurements of real traffic \cite{HBG86}. One expects a transition
from laminar flow to start-stop waves with increasing car density.
For $\vm=1$ it is easy to see that the fundamental diagram is
symmetric with respect to $\rho=1/2$ due to particle-hole symmetry.
This is not true for realistic traffic where on finds a distinct
asymmetry where the maximum of the flow is shifted to lower values of
$\rho$ ($\sim 0.2$).

In the simple mean-field theory approach on assumes that two
neighbouring sites on the ring are completely uncorrelated. Instead
of applying the four update-rules in the order 1-2-3-4 we use the
order 2-3-4-1, i.e.\ we write down the evolution equations of the
configurations of the system after the first step. This has the
advantage that every site is in one of the $\vm+1$ states
$v=0,1,\ldots,\vm$ where a state $v\ge 1$ denotes a car with velocity
$v$ and the state 0 denotes an empty site.  Note, that there are no
cars with velocity zero since after the acceleration step every car
has at least velocity 1. The technical advantage of this procedure
lies in the fact that one reduces the number of evolution equations
by 1 without changing the result. The changed ordering 2-3-4-1 has to
be taken into account in the calculation of the flow $q$. Therefore
the complete dynamics of the systems are determined by a set of $\vm$
coupled nonlinear equations for the densities $c_v(t)$ of cars with
velocity $v$. In general, this time-dependent equations cannot be
solved exactly due to the nonlinearities. However, the equilibrium
properties in the mean-field limit can be derived exactly for any
finite velocity $\vm$ \cite{INSS93} (see Fig.~1 for $\vm=1$ and
Fig.~3 for $\vm=2$).

In the next step we improve the simple mean-field theory by taking
into account neighbour-correlations. In the $n$-site approximation
one writes down self-consistent evolution equations for chain
segments of length $n$ where neighbouring segments have $n-1$ sites
in common\footnote{This $n$-site approximation is similar to the
$(n,n-1)$-cluster approximation of
\cite{BEN92}.}. Here self-consistency means that one has to deal with
conditional probabilities for the sites not belonging to the chain segment
under consideration. The number of equations is $(\vm+1)^n$. Since a car
can drive at most $\vm$ sites in one update-step one should take at least
$n=\vm$ to get reasonable results.

For $\vm=1$ and parallel update the $n$-site approximations with
$n\ge 2$ become identical showing that the 2-site approximation gives
the exact result.  This new result comes not surprising since for
random-sequential update mean-field (1-site approximation) becomes
exact but parallel update rules in general produce stronger
correlations. Explicitely we find in the thermodynamic limit for the
probabilities $P(\sigma_1,\sigma_2)$ to find neighboured sites in the
states $\sigma_1$ and $\sigma_2$ (where $\sigma_j=0,1$ and
$\sigma_j=0$ again denotes an empty site and ${\bar{p}}=1-p$)
\begin{eqnarray}
P(0,1)=P(1,0)&=&{1-\sqrt{1-4{\bar{p}}\rho(1-\rho)}\over 2{\bar{p}}},
\nonumber \\
P(0,0)&=&1-\rho-P(1,0),\nonumber\\
P(1,1)&=&\rho-P(1,0).\nonumber\\
\label{prob}
\end{eqnarray}
The corresponding 'groundstate' for a finite system is given by
\begin{equation}
{\cal P}(N,L)={1\over {\cal{N}}}\sum_{\{\sigma\}}\,\hskip-4pt ' \prod_{j=1}^L
P(\sigma_j,\sigma_{j+1}).
\label{ground}
\end{equation}
Here ${\cal{N}}$ is a normalization constant and the sum $\sum '$
runs over all configurations with a fixed number $N$ of cars
(i.e.~$\sum_{j=1}^L\sigma_j=1$). This result can be shown to be exact for
any finite ring with a fixed number of cars. It is interesting to
note that in contrast to random-sequential dynamics parallel dynamics
lead to an effective attraction between 'particles' and 'holes' (i.e.
$P(0)P(1)=\rho (1-\rho)
\le P(01)$) and thus to a higher flow. This flow
can obtained from the probabilities (\ref{prob}) and yields the following
fundamental diagram (see Fig.~2)
\begin{equation}
q=(1-p)P(1,0)={1-\sqrt{1-4{\bar{p}}\rho(1-\rho)}\over 2}.
\label{flux}
\end{equation}

For $\vm=2$ we have investigated the $n$-site approximations up to
$n=5$.  Unfortunately, the approximation seems not to become exact in
this case.  As can be seen from Fig.~3 the maximum difference between
the 4- and 5-site approximation is less than 1\%. This suggests that
these approximations are already close to the exact result. This
assertion is further supported by the excellent agreement with the
numerical simulations \cite{INSS93}. As can be seen from Fig.~3 the
neighbour correlations lead to an increase of the flux $q$.  In
addition we like to point out that the fundamental diagram for $\vm
=2$ shows the asymmetry known from experimental data \cite{HBG86}.
The details of the method and further results will be published
elsewhere \cite{INSS93,SS93}.

Apart from periodic boundary conditions also other types of boundary
conditions are relevant for the description of traffic flow. Indeed,
it is very important for measurement of real traffic whether the
observed situation corresponds to free traffic or to part of a
bottleneck where the long-range correlations are stronger
\cite{HBG86}. The most natural situation for a bottleneck occurs if
one considers the reduction (over a finite length) of two-lane to
one-lane traffic due to obstacles on one lane. At the end one has
again two-lane-traffic. In principle, this is equivalent to one-lane
traffic with given input- and output-rate at the boundaries. From a
technical point of view the bottleneck situation is - even in
mean-field approximation - much more complicated.  Whereas in the
ring geometry the density is fixed by the initial condition in the
bottleneck the system evolves itself to a stationary state with a
certain mean density. This can be seen explicitly for $\vm =1$ and
random sequential dynamics in \cite{DDM92}-\cite{SCHD93}
where the flux through the bottleneck is
independent of the input- and output-rates over a wide range of these
parameters. For larger velocities $\vm >1$ the bottleneck situation will be
studied in a future publication \cite{SS93}.

We also investigated simple models for two-lane traffic. Here the
update rules are not defined by 1-4. Instead we have introduced fixed
densities $\rho_1$ and $\rho_2$ of cars with velocity 1 and 2,
respectively. Surprisingly we find in mean-field approximation that
the fundamental diagram of each lane is asymmetric but the maximum is
shifted to larger values of $\rho$ ($\rho_{max} > 1/2$). Further
investigations will show if this is an artefact of mean-field theory
or an intrinsic property of our model \cite{INSS93}.

In conclusion, we have studied simple automaton models for the
description of traffic flow. It seems that the equilibriium
properties of these types of models can be described accurately by
improved mean-field theories ($n$-site approximations). On the other
hand the dynamics of these systems seems to be rather complicated. We
like to stress that these models also can be used to describe real
physical problems such as asymmetric exclusion processes and surface
roughening. We also have applied the improved mean-field theory
\cite{SS93} to the Domany-Kinzel cellular automaton \cite{DK} and
find a phase diagram which is in very good agreement with recent
computer simulation studies \cite{tsallis,KS92}.
\\ \\ \\
{\bf Acknowledgement} \\
This work has been performed within the research program of the
Sonderforschungsbereich 341 (K\"oln-Aachen-J\"ulich).\\
We like to thank N.~Ito and K.~Nagel for useful discussions.

\newpage\noindent
{\Large{{\bf Figure Captions}}}
\bigskip
\begin{enumerate}
\item[Fig.~1:] {Fundamental diagram as obtained from the mean-field theory
(dotted line) and from the 2-site approximation for $v_{max}=1$ and
$p={\bar{p}}=1/2$.  Note that the 2-site approximation already
gives the exact result.}
\item[Fig.~2:] {Exact fundamental diagram for $v_{max}=1$ and different values
of $p$.}
\item[Fig.~3:] {Fundamental diagram for $v_{max}=2$ and $p=1/2$ as
obtained from the 1-site approximation (lowest curve) up to the 5-site
approximation (highest curve).}
\end{enumerate}

\end{document}